\documentclass[aps,pre,twocolumn,amsmath,amssymb,superscriptaddress,floatfix,longbibliography]{revtex4-2}

\usepackage{graphicx}
\usepackage{bm}
\usepackage[usenames]{color}
\bibstyle{apsrev.bib}

\newcommand{\be}{\begin{equation}}
\newcommand{\ee}{\end{equation}}
\newcommand{\beqn}{\begin{eqnarray}}
\newcommand{\eeqn}{\end{eqnarray}}

\begin{document}

\title{Renormalization theory of disordered contact processes with heavy-tailed dispersal}

\author{R\'obert Juh\'asz}
\email{juhasz.robert@wigner.hu}
\affiliation{Wigner Research Centre for Physics, Institute for Solid State
Physics and Optics, H-1525 Budapest, P.O.Box 49, Hungary}
     
\date{\today}

\begin{abstract}
Motivated by long-range dispersal in ecological systems, we formulate and apply a general strong-disorder renormalization group (SDRG) framework to describe one-dimensional disordered contact processes with heavy-tailed, such as power law, stretched exponential, and log-normal dispersal kernels, widely used in ecology.
The focus is on the close-to-critical scaling of the order parameters, including the commonly used density, as well as the less known persistence, which is non-zero in the inactive phase. Our analytic and numerical results obtained by SDRG schemes at different levels of approximation reveal that the more slowly decaying dispersal kernels lead to faster-vanishing densities as the critical point is approached. The persistence, however, shows an opposite tendency: the broadening of the dispersal makes its decline sharper at the critical point, becoming discontinuous for the extreme case of power-law dispersal. The SDRG schemes presented here also describe the quantum phase transition of random transverse-field Ising chains with ferromagnetic long-range interactions, the density corresponding to the magnetization of that model.           
\end{abstract}

\maketitle

\section{Introduction}

The contact process (CP) \cite{cp,liggett} is a stochastic lattice model with a widespread use in epidemic spreading and population dynamics. 
It consists of two kinds of competing local processes running on binary state variables attached to each site: Active sites can either spontaneously become inactive or activate nearby inactive sites.  In the context of population dynamics these processes can be interpreted as the extinction of a local population at a habitat patch represented by the sites of the lattice and the colonization of empty habitat patches, respectively. From the side of statistical physics the interest in this model is supplied by its nonequilibrium (absorbing) phase transition \cite{md}, which falls into the universality class of directed percolation (DP) \cite{hhl,odor}, and can be interpreted as an extinction transition in the context of population dynamics.
Although the CP in its simplest form, i.e. with uniform transition rates and colonization of nearest-neighbor sites only, gives correctly an account of the extinction transition, it is inadequate for the purpose of modelling real populations for at least two reasons. First, the conditions of living and reproduction may not be uniform in habitat patches, i.e. the environment is heterogeneous. This can be taken into account in the CP by considering random, site dependent colonization and extinction rates. This kind of quenched disorder in the CP has been thoroughly studied by the strong-disorder renormalization group (SDRG) method \cite{hiv,im} as well as by Monte Carlo simulations \cite{moreira,vd,vfm}, revealing a striking impact in low dimensions not only on the critical behavior of CP but also on the off-critical dynamics. 
The former, namely, is controlled, at least for sufficiently strong disorder \cite{wd_footnote}, by a so-called infinite-disorder fixed point (IDFP) \cite{fisher,hiv,im}, at which dynamical scaling relations involve the logarithm of time rather than the time itself \cite{moreira}. Here, the critical behavior is universal, i.e. independent of the form of disorder. 
The off-critical relaxation is characterized by non-universal power laws \cite{noest} analogous to Griffiths-McCoy singularities of random quantum magnets \cite{griffiths}.
Second, there is a large body of observations in ecology about dispersal processes especially for the pollen or seed dispersal of various plant species \cite{mollison,nathan,nathan_rev,bullock,petrovskii,reynolds,hirsch}. The most relevant characteristic of this process is the so-called dispersal kernel \cite{nathan_rev}, which is the probability density of dispersal distance. According to measurements, typically it has a heavy tail, i.e. it is not confined by an exponential function. As fitting functions to measured dispersal kernels as well as in theoretical modeling various probability density functions are used, although the underlying mechanism leading to a particular heavy-tailed dispersal kernel is in general not clarified. Focusing on heavy-tailed ones, the probability densities widely used in ecology literature can be categorized into three classes concerning their tails at large distances $l$.  These are the power law (PL), $p(l)\sim l^{-\alpha}$, with $\alpha>0$, the stretched exponential (SE), $p(l)\sim e^{-{\rm const}\cdot l^a}$, with $0<a<1$, and the log-normal (LN), $p(l)\sim e^{-{\rm const}\cdot (\ln l)^2}$  probability density functions. 
In the homogeneous CP with a PL dispersal kernel, field-theoretical renormalization group \cite{janssen} and numerical simulations \cite{howard} revealed the following scenario of the critical behavior \cite{hinrichsen}, which is common also for phase transitions in long-range equilibrium systems like the Ising and O(N) models \cite{fmn,sak,bloete,picco,parisi}. For $\alpha>\alpha_{\rm SR}(d)$, where $\alpha_{\rm SR}(d)$ is some dimension dependent threshold, the long-range dispersal is irrelevant, and the model remains in the short-range DP universality class. For $\alpha_{\rm MF}(d)<\alpha<\alpha_{\rm DP}(d)$, the critical exponents vary continuously with $\alpha$, whereas for $\alpha<\alpha_{\rm MF}(d)$, where $\alpha_{\rm MF}(d)=\frac{3}{2}d$ for $d<4$, the critical behavior obeys mean-field theory. 
This means that among the types of dispersal kernels used in ecology only the PL type is able to change the universality class of the extinction transition; the other two (SE and LN) are irrelevant in this respect.      
This is, however, not the case for the disordered CP. 
Recent SDRG studies have revealed that the disordered CP with a PL dispersal kernel has a finite-disorder fixed point (FDFP) for any value of the exponent $\alpha$ in the extensive regime $\alpha>d$ \cite{jki,2dlr,3dlr}. Here, as it has also been confirmed by Monte Carlo simulations, at least in the non-mean-field regime $\alpha>\frac{3}{2}d$ of the homogeneous model, where weak disorder is relevant according to Harris criterion \cite{abh,noest,2dlr}, the logarithmic dynamical scaling characteristic of an IDFP is replaced by power laws, although with non-trivial corrections. 
Furthermore, even a SE type of dispersal kernel has been shown to be able to change the IDFP of the short-range model to a different type of long-range IDFP \cite{stretched,lrpers}. This occurs if $a<\psi_{\rm SR}$, where $\psi_{\rm SR}$ is the exponent appearing in the dynamical relationship $\ln\tau\sim\xi^{\psi_{\rm SR}}$ between temporal and spatial correlation lengths of the corresponding short-range model.  

In this paper, we consider one-dimensional disordered contact processes with different types of heavy-tailed dispersal kernels. We provide a general recipe for constructing an asymptotic SDRG theory which describes the large-scale behavior of the model for a general form of heavy-tailed dispersal kernel. We then revisit the PL and SE classes studied earlier, as well as the enhanced power-law tail, $p(l)\sim e^{-{\rm const}\cdot (\ln l)^{\alpha}}$ with $\alpha>1$, a generalization of the LN tail, which has not been considered so far. We complete earlier studies on PL and SE dispersal kernels by investigating the active phase close to the critical point and studying the vanishing of the stationary density. As an alternative order parameter we also discuss the behavior of the persistence probability in the inactive phase \cite{pers}.  We find that the functional forms appearing in critical scaling relations, including that of the vanishing of the order parameters are determined by the tail of the dispersal kernel. Besides the asymptotic theory, we also study more complete variants of the SDRG method numerically, and compare it with the mainly analytic results of the asymptotic SDRG theory.
Our results show that the heavier tail the dispersal kernel has the more rapidly the order parameter tends to zero on approaching the extinction threshold. 
Interestingly, the persistence follows an opposite tendency: broader dispersal kernels make its transition sharper. 
The former feature may be relevant for ecological modeling in the presence of an environmental gradient \cite{oborny}, where the dependence of the density on the control parameter transforms to an explicit coordinate dependence along the gradient direction.    
  
The paper is organized as follows. In section \ref{sec:sdrg}, SDRG treatments at different levels of approximation are formulated for the CP with a general form of heavy-tailed dispersal kernel. The way of calculating the density and persistence order parameters within the SDRG approach is also outlined. The detailed derivation of various master equations are presented in Appendices \ref{app:master} and \ref{app:order}. In section \ref{sec:analysis}, the machinery of renormalization developed in the previous section is applied to particular forms of dispersal kernels with a focus on the close-to-critical scaling of order parameters. The analysis is performed mainly at the highest level of approximation but lower level numerical schemes are also applied. Finally, the results obtained for the order parameters are discussed in section \ref{discussion}.

\section{The SDRG approach of the contact process} 
\label{sec:sdrg}

The quenched disordered contact process is a continuous-time Markov process with two kinds of local transitions, which occur randomly and independently. Active sites become inactive with site-dependent, quenched rates $\mu_n$, which are drawn independently from some yet unspecified distribution. Active sites can also activate other inactive sites, and we assume for the sake of simplicity that the attempt rate $\lambda$ of this process depends only on the distance $l$ between the source and target sites. The function $\lambda(l)$ tends to zero in the limit $l\to\infty$; otherwise its functional form is kept general at this point. It is, up to a global factor, which can be used as a control parameter of the extinction transition, nothing but the dispersal kernel. For technical reasons (to avoid unambiguity in the order of decimations) we also assume that the sites are located on a line randomly, so that the distance between neighboring sites is an independent, continuous (quenched) random variable. We assume, furthermore, that the large-$l$ tail of the distribution of $l$ is upper bounded by an exponential function.

\subsection{The full SDRG scheme}

The SDRG method for the one-dimensional CP with nearest-neighbor dispersal was formulated and analyzed in Ref. \cite{hiv}. By this procedure, blocks of sites containing the largest transition rate are consecutively replaced by smaller blocks, thereby gradually reducing the number of degrees of freedom, as well as the rate scale $\Omega$ which is set by the actually largest rate. 
If the largest rate is an activation rate, $\Omega=\lambda_{n,n+1}$, provided that the adjacent deactivation rates are much smaller, $\mu_n, \mu_{n+1}\ll\Omega$, the sites $n$ and $n+1$ are clustered and treated as a single degree of freedom. Its effective deactivation rate is obtained perturbatively in leading order as 
\be
\tilde\mu=\kappa\frac{\mu_n\mu_{n+1}}{\Omega},
\label{mu_rule}
\ee
with $\kappa=2$. 
If the largest rate is a deactivation rate, $\Omega=\mu_n$, and $\lambda_{n-1,n}, \lambda_{n,n+1}\ll\Omega$, then site $n$, being almost always inactive, is eliminated, leaving behind a direct activation rate between sites $n-1$ and $n+1$. This is obtained again by perturbation calculation in leading order as 
\be
\tilde\lambda_{n-1,n+1}=\frac{\lambda_{n-1,n}\lambda_{n,n+1}}{\Omega}.
\label{sr_rule}
\ee 

\subsection{Approximative nearest-neighbor schemes} 

The difficulty about long-range dispersal within the SDRG method is that, owing to the all-to-all connection, the elimination of a site renormalizes all remaining transition rates. In order to keep the model of one-dimensional structure and thereby analytically tractable, simplified SDRG schemes were introduced in Ref. \cite{jki,stretched}.

\subsubsection{The first nearest-neighbor (NN1) scheme}

The first step in a series of approximations is that the long-range interaction (activation) between clusters is taken into account when, in the course of the SDRG procedure, they become directly adjacent. In other words, at any stage of the procedure, only the interactions between neighboring clusters (being the most relevant) are kept, while those between farther neighbors are dropped. Thereby the one-dimensional structure is restored, and to distinguish this approximation from further ones, we will call it the first nearest-neighbor (NN1) scheme.        
Here, if the largest rate is an activation rate between clusters $\mathcal{C}_n$ and $\mathcal{C}_{n+1}$,  $\Omega=\lambda_{n,n+1}$, a new cluster $\mathcal{C}_{\tilde n}=\mathcal{C}_n\cup\mathcal{C}_{n+1}$ is formed with an effective deactivation rate given in Eq. (\ref{mu_rule}). At the same time, the activation rate between $\mathcal{C}_{\tilde n}$ and $\mathcal{C}_{n+2}$ is modified to
\be 
\lambda_{\tilde n,n+2}=\lambda_{n+1,n+2}+\sum_{i\in \mathcal{C}_n, j\in \mathcal{C}_{n+2}}\lambda(l_{ij}),
\ee
and, similarly, the activation rate to $\mathcal{C}_{n-1}$ will be $\lambda_{n-1,\tilde n}=\lambda_{n-1,n}+\sum_{i\in \mathcal{C}_{n-1}, j\in \mathcal{C}_{n+1}}\lambda(l_{ij})$.
If the largest rate is $\Omega=\mu_n$, cluster $\mathcal{C}_{n}$ is deleted and the new activation rate between cluster $\mathcal{C}_{n-1}$ and $\mathcal{C}_{n+1}$ will be 
\be
\tilde \lambda_{n-1,n+1}=\frac{\lambda_{n-1,n}\lambda_{n,n+1}}{\Omega} + 
\sum_{i\in \mathcal{C}_{n-1}, j\in \mathcal{C}_{n+1}}\lambda(l_{ij}).
\ee
This scheme was applied numerically for the SE dispersal kernel in Ref. \cite{stretched}.

\subsubsection{The second nearest-neighbor (NN2) scheme}

The next step toward analytic tractability is that the activation rate between neighboring clusters is approximated by the long-range activation rate between the closest constituents of the clusters, which is $\lambda(l_{nm})$, where $l_{nm}$ denotes the distance between them. This approximation becomes more accurate for more rapidly decreasing dispersal kernels; for the worst case, the PL dispersal kernel, the error made by this approximation has been estimated a posteriori,  showing that it affects the power of multiplicative logarithmic corrections to dynamical scaling relations \cite{jki,2dlr}. 
This level of approximation will be called the second nearest-neighbor (NN2) scheme. Here, the renormalized system is described by three sets of parameters: the activation rates $\lambda(l_{n,n+1})$, or equivalently the distances $l_{n,n+1}$ between adjacent clusters, and the deactivation rate $\mu_n$ and width $w_n$ of clusters. In the case $\Omega=\lambda(l_{n,n+1})$, the rule in Eq. (\ref{mu_rule}) is then extended with the transformation of lengths
\be 
\tilde w=w_n+l_{n,n+1}+w_{n+1},
\label{w_rule}
\ee     
while, for $\Omega=\mu_{n}$ we have simply
\be
\tilde l=l_{n-1,n}+w_n+l_{n,n+1}.
\label{l_rule}
\ee

\subsubsection{The third nearest-neighbor (NN3) scheme}
\label{sec:nn3}

The next approximation can be performed only if the $\mu$-decimation events are much more frequent than the $\lambda$-decimations, which is valid in the following cases. First, in the inactive phase and at the critical point, at late stages of the renormalization. Second, it is also valid in the active phase, close to the critical point, at late stages but only until the renormalization trajectory (see later) is close to the critical one. The SE model is an exceptional case, as will be discussed later in section \ref{sec:se}. Here, the NN3 scheme is invalid in the active phase and at the critical point and applicable only in the inactive phase.
Under the above restrictions, the width $w_n$ of clusters will be small compared to the spacings $l_n$ between them and can be neglected. This means that the variables $w_n$ together with the rule in Eq. (\ref{w_rule}) are dropped, and we are left with two sets of variables, $l_{n,n+1}$ and $\mu_n$. Decimation of a $\lambda$ rate is described by Eq. (\ref{mu_rule}) as before, whereas, in the case of a $\mu$-decimation, Eq. (\ref{l_rule}) reduces to 
\be
\tilde l=l_{n-1,n}+l_{n,n+1}.
\label{l_rule2}
\ee 
This renormalization scheme will be called the third nearest-neighbor (NN3) scheme.

\subsubsection{Summary of SDRG schemes}

The different approximations involved in the hierarchy of SDRG schemes presented so far for the CP with long-range dispersal can be summarized as follows. In the full SDRG method, all interactions between the clusters are taken into account at any stage of the renormalization. In the NN1 scheme, only the interactions between adjacent clusters are kept. In addition to this, in the NN2 scheme, the interaction between adjacent clusters is approximated by the interaction between their closest constituent sites and the contributions of other pairs of sites are dropped. Finally, in addition to all these approximations, the spatial extension of clusters is also neglected in the NN3 scheme. This latter scheme is valid  only with the limitations described in section \ref{sec:nn3}.

\subsection{Handling of the problem about $\kappa>1$}

Before analyzing the NN2 and NN3 schemes, we mention that the above SDRG schemes with the only modification $\kappa=1$ in Eq. (\ref{mu_rule}) describe the random transverse-field Ising chain with long-range ferromagnetic couplings $\lambda(l)$ \cite{jki}. A $\kappa$ parameter exceeding $1$, as in the case of the CP, makes a further complication of the method. The generated rate $\tilde\mu$, namely,  may happen to be greater than $\Omega$, making the variation of $\Omega$ non-monotonic in the course of the renormalization. As these events are of vanishing probability when an IDFP is approached, the simplest way of circumvent this problem is to choose $\kappa=1$ \cite{sm}. This is, however, not justified for the PL dispersal kernel, which is described by an FDFP rather than an IDFP \cite{jki}. Nevertheless, as it was shown in Ref. \cite{pers}, the case $\kappa=2$ can also be analytically treated within the NN3 scheme of the PL and SE dispersal kernels, leading to modified flow equations compared to the $\kappa=1$ case. 
The key point of this treatment is that, a generated rate $\tilde\mu$ for which $\tilde\mu>\Omega$ is immediately decimated by a $\mu$-decimation step. This two-step composite decimation will be referred to as an anomalous $\lambda$-decimation, to distinguish it from normal $\lambda$-decimations for which $\tilde\mu<\Omega$. In deriving the NN3 scheme for general forms of dispersal kernels we will therefore not restrict ourselves to $\kappa=1$.

Concerning these anomalous $\lambda$-decimations, one could object that the perturbative decimation rule in Eq. (\ref{mu_rule}) containing the prefactor $\kappa$ is correct only if $\mu_n, \mu_{n+1}\ll\Omega$, whereas for anomalous decimations, for which $\mu_n, \mu_{n+1}\sim\Omega$, it is not justified.
But to construct an analytically tractable scheme one needs to use a uniform decimation rule, i.e. that in Eq. (\ref{mu_rule}) with a constant prefactor $\kappa$ irrespective of how good the conditions of perturbative treatment are fulfilled. In such a scheme, which contains the asymptotically correct, constant prefactor $\kappa$, the handling of (rare) anomalous decimations (although these are strongly approximative) is technically necessary.

\subsection{Master equation for rate distributions}

We have seen in the previous section that, in the NN3 scheme, we have two sets of variables: $\{l_n\}$ which are perfectly correlated with the rates $\lambda_n$ through the function $\lambda(l)$ and the deactivation rates $\{\mu_n\}$. 
If these variables are independent, random variables in the initial model, they remain so in the course of the SDRG procedure, therefore it is sufficient to characterize the renormalized system at some rate scale $\Omega$ by the distributions $G_{\Omega}(\mu)$ and $F_{\Omega}(l)$. When $\Omega$ changes, these distribution also change and one can derive a master equation governing their evolution during the SDRG procedure. The details are presented in Appendix \ref{app:master}. 
For the sake of simplicity, we will assume in the following analytic treatment that both $\lambda_n$ and $\mu_n$ are distributed initially in the same range $(0,\Omega_0)$. Later, in the numerical analysis, this restriction will be relaxed.
Rather than using the original variables $\Omega$, $\mu$, and $l$, it is expedient to use the logarithmic rate scale 
\be
\Gamma=\ln(\lambda_0/\Omega),
\label{Gamma}
\ee
where $\lambda_0$ is a constant rate, appearing in the dispersal kernel,
and the following reduced variables: 
\be 
\beta=\ln(\Omega/\mu),
\label{beta}
\ee
and 
\be
\zeta=\frac{l}{\lambda^{-1}(\Omega)}-1.
\label{zeta}
\ee
In the latter, $\lambda^{-1}$ denotes the inverse function of the dispersal kernel, so $\lambda^{-1}(\Omega)$ is just the lower edge of the distribution of spacings $l$ between clusters at scale $\Omega$. This is the point where the form of the dispersal kernel enters the problem and for technical reasons we also introduce the characteristic function of the dispersal kernel 
\be
\Theta(\Gamma)=\ln[\lambda^{-1}(\Omega)]
\label{theta}
\ee
and its derivative $\Theta^{\prime}\equiv\frac{d\Theta}{d\Gamma}$.
In terms of these new variables, the rule of $\lambda$-decimations in Eq. (\ref{mu_rule}) transforms to 
\be
\tilde\beta=\beta_n+\beta_{n+1}-B,
\label{beta_rule}
\ee
where $B=\ln\kappa=\ln 2$, while the rule of $\mu$-decimations in Eq. (\ref{l_rule2}) can be written as $\tilde\zeta=\zeta_{n-1,n}+\zeta_{n,n+1}+1$. 
A complete treatment of latter rule with the additive positive constant $1$ is difficult, as it leads to nonanalyticity in the distribution of $\zeta$, see Ref. \cite{juhasz2009} and references therein. 
Nevertheless, we will see that the distribution of $\zeta$ variables is broadening in the course of the renormalization for all cases in the domain of validity of the NN3 scheme, therefore we may drop the constant term and write 
\be
\tilde\zeta=\zeta_{n-1,n}+\zeta_{n,n+1}.
\label{zeta_rule}
\ee
As it is derived in Appendix \ref{app:master}, the decimation rules in Eqs. (\ref{beta_rule}) and (\ref{zeta_rule}) lead to the following master equations in terms of the probability densities $g_{\Gamma}(\beta)$ and $f_{\Gamma}(\zeta)$:
\begin{widetext}
\beqn 
\frac{\partial g_{\Gamma}(\beta)}{\partial\Gamma}&=& 
\frac{\partial g_{\Gamma}(\beta)}{\partial\beta} +  g_{\Gamma}(\beta)[g_0-f_0\Theta^{\prime}p_{\Gamma}] +f_0\Theta^{\prime}\int_{0}^{\beta+B} g_{\Gamma}(\beta') g_{\Gamma}(\beta-\beta'+B)d\beta'  \label{master_g} \\
\frac{\partial f_{\Gamma}(\zeta)}{\partial\Gamma}&=& 
\Theta'(\zeta+1)\frac{\partial f_{\Gamma}(\zeta)}{\partial\zeta} + f_{\Gamma}(\zeta)[f_0\Theta'p_{\Gamma}-g_0+\Theta'] + 
[g_0+\Theta'f_0(1-p_{\Gamma})]\int_0^{\zeta}f_{\Gamma}(\zeta')f_{\Gamma}(\zeta-\zeta')d\zeta', \label{master_f}
 \eeqn
\end{widetext}
where $g_0(\Gamma)\equiv g_{\Gamma}(0)$, $f_0(\Gamma)\equiv f_{\Gamma}(0)$, and $p_{\Gamma}$ is the probability of normal $\lambda$-decimations (for which $\tilde\mu<\Omega$), see Appendix \ref{app:master} for details. 
These equations have a one-parameter solution of simple form:
\beqn
g_{\Gamma}(\beta)=g_0e^{-g_0\beta}  \label{sol_g}  \\
f_{\Gamma}(\zeta)=f_0e^{-f_0\zeta}, \label{sol_f}
\eeqn  
in which the dependence on $\Gamma$ enters through the functions $g_0(\Gamma)$ and $f_0(\Gamma)$.
Using this, the probability of normal $\lambda$-decimations can readily be evaluated to yield
\be 
p_{\Gamma}=e^{-g_0B}(1+g_0B). 
\ee
Substituting Eqs. (\ref{sol_g}-\ref{sol_f}) into the master equations, we obtain the following flow equations for $g_0$ and $f_0$: 
\beqn
g_0'&=&-\Theta'f_0g_0e^{-g_0B} \label{flow_g}  \\
f_0'&=&-f_0[g_0-\Theta'+f_0\Theta'(1-p_{\Gamma})]. \label{flow_f}
\eeqn
We stress, however, that Eq. (\ref{sol_f}) and Eq. (\ref{flow_f}) are not valid for the SE dispersal kernel outside of the inactive phase.

\subsection{Order parameters}

Having derived the flow equations for parameters characterizing the distribution of rates, we turn to the question how the dynamical and stationary properties of the model can be inferred from the SDRG solution. 
First, we introduce the ratio $r$ of the frequencies of $\lambda$-decimations to $\mu$-decimation, which is given by 
\be 
r=\frac{f_0\Theta'}{g_0+f_0\Theta'(1-p_{\Gamma})}.
\label{r}
\ee
It tends to zero (infinity) in the inactive (active) phase, and it is therefore  useful for locating the critical point in numerical SDRG analyses. At the critical point it decays to zero in a way that depends on the concrete form of the dispersal kernel. 

The relationship between the time scale $t=1/\Omega$ and the length scale $l=1/n$, where $n$ is the fraction of sites not yet decimated up to scale $\Omega$, is also related directly to $g_0$ and $f_0$. 
By an infinitesimal change of $\Gamma$, $n$ changes as 
\be
\frac{dn}{n}=-[g_0+f_0\Theta'(2-p_{\Gamma})]d\Gamma,
\ee
which leads, by integration, to the relationship  
\be 
l=l_0e^{\int[g_0+f_0\Theta'(2-p_{\Gamma})]d\Gamma}.
\label{dyn}
\ee

\subsubsection{Density of active sites}

A widely used order parameter of the phase transition of the CP is the global density of active sites in the stationary state \cite{md}. More generally, one is interested in the dependence of the global density on time when the process is initiated from a non-stationary state, most frequently from a fully active state. Within the SDRG theory, a given site is active at time $t$ if it has not been eliminated yet by a $\mu$-decimation event until scale $\Omega=1/t$; otherwise it is inactive. The global density is thus given by the survival probability $\rho_b(\Gamma)$ of sites under the SDRG procedure in the above sense. Besides bulk sites in an infinite system, we can also consider the local density at the first (surface) site of semi-infinite system, which is thus given by the survival probability $\rho_s(\Gamma)$ of the first site. 
Flow equations for $\rho_b(\Gamma)$ and $\rho_s(\Gamma)$ can be obtained by using the decompositions $\rho_b(\Gamma)=\int s^{(b)}_{\Gamma}(\beta)d\beta$ and $\rho_s(\Gamma)=\int s^{(s)}_{\Gamma}(\beta)d\beta$, where the integrands are the probabilities that a given bulk or surface site, respectively, survived the $\mu$-decimations up to scale $\Gamma$ in a cluster having a $\beta$ variable in the range $[\beta,\beta+d\beta]$. As it is derived in Appendix \ref{app:order}, they obey the following master equations 
\begin{widetext}
\be
\frac{\partial s^{(i)}_{\Gamma}(\beta)}{\partial\Gamma}= 
\frac{\partial s^{(i)}_{\Gamma}(\beta)}{\partial\beta} -
n_if_0\Theta^{\prime}\left[s^{(i)}_{\Gamma}(\beta)-\int_{0}^{\beta+B} s^{(i)}_{\Gamma}(\beta') g_{\Gamma}(\beta-\beta'+B)d\beta'\right],  \label{master_s} 
\ee
with $i=b,s$ and $n_s=1$, $n_b=2$.
\end{widetext}
The solutions of these equations are of the form 
\beqn
 s^{(s)}_{\Gamma}(\beta)&=&\rho_s(\Gamma)g_0e^{-g_0\beta} \label{trial_s} \\
 s^{(b)}_{\Gamma}(\beta)&=&[u(\Gamma)+v(\Gamma)g_0\beta]g_0e^{-g_0\beta},
\label{trial_b}
\eeqn
containing the unknown functions $u(\Gamma)$ and $v(\Gamma)$, which have the initial values $u(\Gamma_0)=1$ and $v(\Gamma_0)=0$, and which add up to the bulk survival probability, $\rho_b(\Gamma)=u(\Gamma)+v(\Gamma)$. 
Substituting Eqs. (\ref{trial_s}-\ref{trial_b}) into the master equations, yields the following flow equations for the surface order parameter 
\be
\rho^{\prime}_s(\Gamma)=-\rho_s(\Gamma)[g_0+f_0\Theta'(1-p_{\Gamma})],
\label{flow_surf}
\ee
and for the components of the bulk order parameter
\beqn
u'&=&u[-g_0-f_0\Theta'e^{-g_0B}+2f_0\Theta'(p_{\Gamma}-1)] + \nonumber \\ 
&+& v[g_0+f_0\Theta'g_0^2e^{-g_0B}B^2] \\
v'&=&uf_0\Theta'e^{-g_0B} + v[-g_0+2f_0\Theta'(p_{\Gamma}-1)]. \nonumber \\ 
\eeqn
Comparing Eq. (\ref{flow_surf}) with Eq. (\ref{flow_f}) we find that
$[\ln(\rho_s)]'=[\ln(f_0)]'-\Theta'$, therefore the surface density is related to the parameter $f_0$ as 
\be
\rho_s(\Gamma)=\frac{f_0(\Gamma)}{f_0(0)}e^{-\Theta(\Gamma)+\Theta(0)}={\rm const}\cdot f_0(\Gamma)e^{-\Theta(\Gamma)}.
\label{Sf}
\ee

\subsubsection{Local persistence}

An alternative order parameter of the CP is the local persistence, which has attracted much attention in the DP universality class \cite{hk,am,menon,fuchs,grassberger,bg,pers,lrpers}. It is defined as the probability that a site which was initially inactive, has not been activated up to time $t$. Here, the density of active sites in the initial state is assumed to be less than one; an alternative is that all but one sites are active initially. The persistence in the stationary state, as opposed to the local density, is non-zero in the inactive phase and zero in the active phase. 
In the SDRG approach, an initially inactive site (the $\mu$ rate of which is set to zero for preventing it from elimination) remains persistent until it is merged into a cluster by a $\lambda$-decimation next to it \cite{pers}. 
Furthermore, within the nearest-neighbor SDRG schemes, a given bulk site can loose its persistence either by a $\lambda$-decimation  on its left-hand-side or on its right-hand-side, and these events occur independently. Consequently, the probability $\pi_b(\Gamma)$ that a bulk site remains persistent up to the scale $\Gamma$ is related to that of the surface site of a semi-infinite system $\pi_s(\Gamma)$ as $\pi_b(\Gamma)=[\pi_s(\Gamma)]^2$. 
Thus, it is sufficient to consider the surface persistence $\pi_s(\Gamma)$.
Setting the $\mu$ rate of the first site to zero, we note that $\pi_s(\Gamma)$
is the probability that the first bond has not been eliminated by a $\lambda$-decimation up to the scale $\Gamma$. It is apparent that $\pi_s(\Gamma)$ is dual to the survival probability $\rho_s(\Gamma)$ of the first site, and its fixed-point value can thus be formally regarded as a ``reversed'' order parameter which becomes non-zero in the inactive phase. Decomposing $\pi_s(\Gamma)$ by the $\zeta$ variable as $\pi_s(\Gamma)=\int q_{\Gamma}(\zeta)d\zeta$, we can write the following master equation for $q_{\Gamma}(\zeta)$ (for details, see Appendix \ref{app:order}):
\begin{widetext}
\be 
\frac{\partial q_{\Gamma}(\zeta)}{\partial\Gamma}= 
\Theta'(1+\zeta)\frac{\partial q_{\Gamma}(\zeta)}{\partial\zeta} 
+ \Theta'q_{\Gamma}(\zeta) -
[g_0+f_0\Theta'(1-p_{\Gamma})]\left[q_{\Gamma}(\zeta)-\int_0^{\zeta}q_{\Gamma}(\zeta')f_{\Gamma}(\zeta-\zeta')d\zeta'\right]. \label{master_q}
\ee
\end{widetext} 
The solution of this equation can be found in the form 
$q_{\Gamma}(\zeta)=\pi_s(\Gamma)f_0e^{-f_0\zeta}$, which leads to the following flow equation for $\pi_s(\Gamma)$: 
\be
\pi_s^{\prime}(\Gamma)=-\pi_s(\Gamma)f_0\Theta'.
\label{flow_Q}
\ee
Comparing this equation with Eq. (\ref{flow_g}), we find that 
$[\ln(\pi_s)]'=[\ln(g_0)]'e^{g_0B}$. This does not provide a strictly linear relationship like the analogous formula in Eq. (\ref{Sf}), nevertheless, in the limit $g_0\to 0$, we still have an asymptotic proportionality, $\pi_s\sim g_0$, the exponential factor giving only corrections to this.

\section{Analysis of the flow equations}
\label{sec:analysis}

Next, we will analyze the SDRG flow equations obtained in the previous section. Although we presented the SDRG description by allowing anomalous $\lambda$-decimations occurring if $\kappa>1$, they turn out to give at most subleading corrections at IDFP-s, while for the PL dispersal kernel, which is described by an FDFP, they modify the coefficient of the leading term of $f_0(\Gamma)$, see Ref. \cite{lrpers} for details. Therefore, we will analyze the flow equations by setting $\kappa=1$, leading to $B=0$ and $p_{\Gamma}=1$, which greatly simplifies the analysis. Note that this is precisely the SDRG theory which describes the random transverse-field Ising model. 
We have thus the following set of flow equations 
\beqn 
g_0'&=&-\Theta'f_0g_0, \label{flow_g_simp}  \\
f_0'&=&-f_0[g_0-\Theta'], \label{flow_f_simp} \\
u'&=&-u[g_0+f_0\Theta'] + vg_0,  \label{u_simp} \\ 
v'&=&uf_0\Theta'  -vg_0, \label{v_simp} 
\eeqn   
where Eq. (\ref{flow_f_simp}) is invalid for the SE dispersal kernel outside of the inactive phase. 
Note that the first two of these equations constitute an autonomous subsystem. 
The bulk and surface (density) order parameter are given by
$\rho_b=u+v$ and $\rho_s={\rm const}\cdot f_0e^{-\Theta}$, respectively, whereas the persistence is $\pi_s={\rm const}\cdot g_0$. 

\subsection{General features}

First, we discuss the properties of the flow diagrams and the scaling of order parameters which are generally valid for all types of heavy-tailed dispersal kernels, then we consider the specialities separately. The flows of the parameters $g_0$ and $f_0$ for the PL, SE, and LN dispersal kernels can be seen in Figs. \ref{fig_pl_flow}, \ref{fig_sefd}, and \ref{fig_ln_flow}. 

Three kinds of trajectories can be distinguished. First, they can end up at some point of the horizontal axis, $g_0={\rm const}$, $f_0=0$. Approaching to such fixed points the decimation ratio in Eq. (\ref{r}) tends to zero; thus, this line of fixed-points describes the inactive phase. From the dynamical relationship in Eq. (\ref{dyn}), we obtain that the limiting value $g_0(\infty)$ is the inverse of the dynamical exponent, $g_0(\infty)\equiv\frac{1}{z}$, which enters in the relationship between length and time scale, $\Omega^{-1}\sim l^z$. This line of fixed points ends at $g_0=0$, except of the PL case, for which it ends at $g_0=\frac{1}{\alpha}$.   

Concerning the density order parameter, we obtain from Eq. (\ref{flow_f_simp}) by setting $g_0=\frac{1}{z}$ at late $\Gamma$ scales that, at the surface, it decays to zero as 
\be
\rho_s(\Gamma)\sim e^{-\frac{1}{z}\Gamma},
\ee
while from Eqs. (\ref{u_simp}-\ref{v_simp}) by neglecting $f_0\Theta'$ we obtain in the bulk the asymptotic decay
\be
\rho_b(\Gamma)\sim \frac{1}{z}\Gamma e^{-\frac{1}{z}\Gamma}.
\ee  
irrespective of the form of the dispersal kernel.  
Setting $\Gamma$ to $\ln(t/t_0)$ in these forms, where $t_0$ is some non-universal constant of time dimension, we obtain an algebraic time dependence of the surface and bulk densities, with a logarithmic correction in the latter case 
\beqn 
\rho_s(t)&\sim& t^{-\frac{1}{z}}, \\
\rho_b(t)&\sim& t^{-\frac{1}{z}}\ln(t/t_0).
\eeqn
This slow decay is caused by rare-region effects \cite{vojta_rev} and is analogous to the Griffiths-McCoy singularities in quantum magnets \cite{griffiths,noest}. 
 
The surface persistence $\pi_s$ in this phase remains non-zero at the fixed points, and tends to zero as the end point of the line of fixed points is approached, except of the PL model, for which it remains non-zero also at this point. 
 
The other class of trajectories, followed by an initial decrease of $f_0$, tend to the point $(g_0=0,f_0=\infty)$. In these cases, the decimation ratio tends to infinity, showing that these trajectories correspond to the active phase of the model. From Eq. (\ref{flow_f_simp}) we can see that $f_0\simeq {\rm const}\cdot e^{\Theta}$, thus the surface density order parameter $\rho_s={\rm const}\cdot f_0e^{-\Theta}$ (just as the bulk one $\rho_b\simeq v$) tends to different non-zero limiting values for each such trajectory. The persistence in this phase tends to zero as $\pi_s={\rm const}\cdot g_0\sim e^{-{\rm const}\cdot e^{\Theta}}$, the form of which thus depending on the dispersal kernel.   

These two classes of trajectories are separated by the critical trajectory ending at the critical fixed point. The behavior of the order parameters at and near this fixed point will be discussed for each type of dispersal kernel separately. 

\subsection{Power-law dispersal kernel}

\begin{figure}[ht]
\includegraphics[width=8cm]{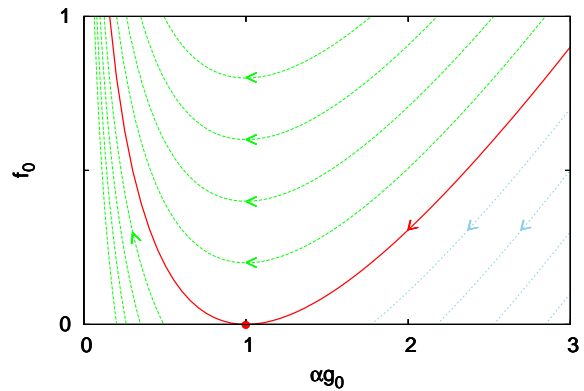}  
\caption{\label{fig_pl_flow} Flow diagram of the CP with PL dispersal kernel by the NN3 scheme. The red solid line is the critical trajectory ending at the critical fixed point $(1,0)$. The blue dotted lines and the green broken lines correspond to the inactive and active phase, respectively.}
\end{figure}

For the PL dispersal kernel, the activation rate decreases with the distance as
\be
\lambda(l)=\lambda_0l^{-\alpha}.
\ee
The characteristic function and its derivative are thus $\Theta=\Gamma/\alpha$ and
\be
\Theta'=\frac{1}{\alpha},
\ee
respectively.
We mention that the flow equations of $g_0$ and $f_0$ for $\alpha=1$ are of Kosterlitz-Thouless type and appear in the SDRG treatment of other models, as well \cite{akpr,narayan,vh_noise}.
The trajectories are given by the equations
\be
f_0=\alpha g_0-\ln(\alpha g_0)-1+\Delta,
\label{tra}
\ee
$\Delta=0$ corresponding to the critical line, see the flow diagram in Fig. \ref{fig_pl_flow}.
Close to the critical fixed point, the $\Gamma$-dependence of the parameters
are $g_0=\frac{1}{\alpha}+2\Gamma^{-1}+O(\Gamma^{-2})$ and $f_0=2\alpha^2\Gamma^{-2}+O(\Gamma^{-3})$. We mention that for the more general case of $\kappa\ge 1$, the latter is modified to $f_0=2\alpha^2\kappa^{1/\alpha}\Gamma^{-2}+O(\Gamma^{-3})$ \cite{lrpers}. The surface and bulk density order parameter for large $\Gamma$ along the critical line are given by
\beqn
\rho_s&\sim& f_0e^{-\Theta}\sim \Gamma^{-2}e^{-\frac{1}{\alpha}\Gamma},
\label{pl_ss_gamma} \\
\rho_b&\sim& e^{-\frac{1}{\alpha}\Gamma},
\label{pl_sb_gamma}
\eeqn
respectively. The latter can be shown not to be affected by the parameter $\kappa$ in the general case $\kappa\ge 1$. The surface persistence at the critical line tends to a non-zero limit as $\pi_s\sim g_0\sim\frac{1}{\alpha}+2\Gamma^{-1}+O(\Gamma^{-2})$. Replacing in all these relations $\Gamma$ with $\ln(t/t_0)$, we obtain the asymptotic dependence of order parameters on time at the critical point (see Table \ref{table}). 

Let us turn to the question how the order parameters behave outside of but close to the critical point. As a reduced control parameter, we can use the deviation $\Delta$ of one of the initial parameters $g_0$ or $f_0$ from the critical line. 
Due to the form of the equations of trajectories, $\Delta$ remains constant during the renormalization. In the inactive phase ($\Delta<0$),  $\pi_s\sim g_0\to {\rm const}=\frac{1}{z}$. Since the critical trajectory is quadratic at the critical point, we have $\frac{1}{z}-\frac{1}{\alpha}= \frac{1}{\alpha}(2|\Delta|)^{1/2}+O(\Delta)$, and the deviation of the persistence from its critical value is thus $O(|\Delta|^{1/2})$.    

In the active phase ($\Delta>0$), the dependence of the density order parameter on $\Delta$ can be obtained by the following argument. From Eqs. (\ref{flow_g_simp}-\ref{flow_f_simp}) we obtain
$\Gamma=\alpha\int[\alpha g_0-\ln(\alpha g_0)-1+\Delta]^{-1}d\ln g_0+{\rm const}$.
Moving along a close-to-critical trajectory, there is a large contribution $\Gamma_c\sim\Delta^{-1/2}$ to this integral at the saddle point $\alpha g_0=1$, beyond which the density order parameter saturates to its limiting value. Up to this crossover scale $\Gamma_c$, the order parameter follows the critical scaling given in Eqs. (\ref{pl_ss_gamma}-\ref{pl_sb_gamma}). Substituting $\Gamma_c\sim\Delta^{-1/2}$ into these equations we obtain
\beqn
\rho_s(\Delta)&\sim& \Delta e^{-\frac{c}{\sqrt{\Delta}}},
\label{pl_ss_delta} \\
\rho_b(\Delta)&\sim& e^{-\frac{c}{\sqrt{\Delta}}},
\label{pl_sb_delta}
\eeqn
as $\Delta\to 0$, with $c$ denoting a non-universal constant. 

\begin{figure}[ht]
\includegraphics[width=8cm]{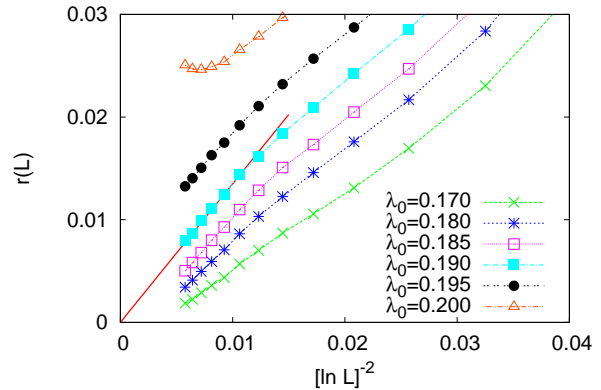}  
\caption{\label{fig_rpl3}  Dependence of the ratio of frequencies of $\lambda$ and $\mu$ decimations on the size, obtained numerically by the NN1 scheme for the PL dispersal kernel with $\alpha=3$. The straight line is a guide to the eye.}
\end{figure}

\begin{figure}[ht]
\includegraphics[width=8cm]{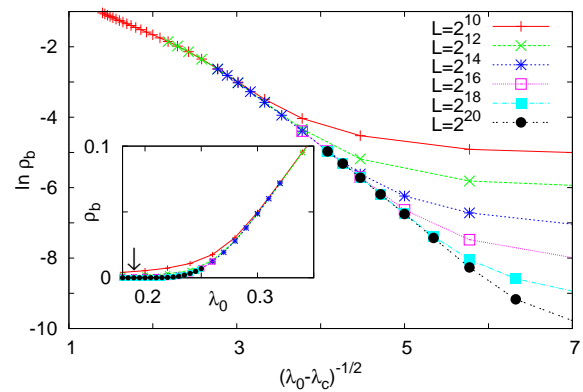}
\caption{\label{fig_oppl3} The bulk density order parameter of the CP with a PL kernel with $\alpha=3$, plotted against the reduced control parameter for different system sizes. The data were obtained by the NN1 scheme. The arrow in the inset indicates $\lambda_c$.}
\end{figure}

Numerical results on the bulk density order parameter obtained with the NN1 scheme and by the numerical integration of the NN3 flow equations (\ref{flow_g_simp}-\ref{v_simp}) are in agreement with Eq. (\ref{pl_sb_delta}). 
We applied the NN1 scheme with $\kappa=1$ and used a uniform distribution of $\mu$ rates in the range $[0,1]$ and equidistant sites, i.e. the nearest-neighbor activation rates were initially $\lambda_0$. Periodic boundary condition was used and the renormalization was carried out up to two clusters, starting with different sizes $L$. 
First, we determined the location of the critical point $\lambda_c$ by considering the size dependence of the decimation ratio $r(L)$ at the last step, which was calculated by performing the renormalization of $10^7$ random samples for each $L$. In the inactive (active) phase $r(L)$ tends to zero (infinity), while at the critical point, it tends to zero, according to Eqs. (\ref{r}) and (\ref{dyn}) in leading order as $r(L)\sim (\ln L)^{-2}$. As can be seen in Fig. (\ref{fig_rpl3}), we obtain in this way the estimate $\lambda_c=0.190(5)$.  
Next, we calculated the bulk density order parameter $\rho_b(L)$ at the final stage of the renormalization by averaging over random samples ($2500$ in number for the largest $L$). The variation of $\rho_b(L)$ with the reduced control parameter $\Delta=\lambda_0-\lambda_c$ for different sizes is shown in Fig. \ref{fig_oppl3}. 
As can be seen in the figure, the saturation value of the bulk density order parameter follows the law in Eq. (\ref{pl_sb_delta}) well.

\subsection{The stretched exponential dispersal kernel}
\label{sec:se}

The SE dispersal kernel is of the form
\be
\lambda(l)=\lambda_0e^{-bl^a}
\ee
with positive constants $a$ and $b$. As it was argued in Ref. \cite{stretched}, this type of dispersal kernel is relevant, i.e. changes the universality class from the short-range one if $a<1/2$. The characteristic function is
$\Theta=-\overline{a}\ln b+\overline{a}\ln\Gamma$, and its derivative is
\be
\Theta'=\overline{a}\Gamma^{-1},
\ee
where we have introduced $\overline{a}\equiv 1/a$.
As said before, a pure exponential distribution of $\zeta$ solves the master equation only in the inactive phase for large $\Gamma$, thus Eq. (\ref{flow_f_simp}) is not valid outside of this phase. Nevertheless, Eq. (\ref{flow_g_simp}) alone fixes the leading term of $g_0$ and $f_0$ at the critical point for large $\Gamma$ \cite{stretched}, which are $g_0\simeq C\Gamma^{-1}$ and $f_0\simeq a$. The constant $C$ can be determined by requiring the dynamical relationship to be $l^a\sim \Gamma$, which is dictated by the form of the dispersal kernel. Using Eq. (\ref{dyn}), we obtain then $C=\overline{a}-1$.
The leading terms of $g_0$ and $f_0$ determine those of the order parameters at the critical point. We obtain for the surface and bulk density order parameters
\beqn
\rho_s(\Gamma)&\sim& \Gamma^{-x_s\overline{a}} \label{rho_se_surf} \\
\rho_b(\Gamma)&\sim& \Gamma^{-x_b\overline{a}}, \label{rho_se_bulk}
\eeqn
respectively, with $x_s\overline{a}=\overline{a}-1$ and $x_b\overline{a}=\overline{a}-\frac{1}{2}-\sqrt{\overline{a}-\frac{3}{4}}$ \cite{stretched}, whereas the surface persistence scales as
\be
\pi_s(\Gamma)\sim \Gamma^{-1}
\ee
at the critical point \cite{lrpers}. 

\begin{figure}[ht]
\includegraphics[width=8cm]{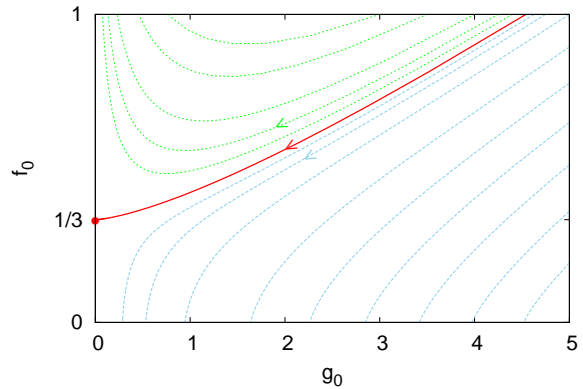}  
\caption{\label{fig_sefd} Hypothetical flow diagram for the SE dispersal kernel with $a=1/3$ obtained by Eqs. (\ref{flow_g_simp}) and (\ref{flow_f_hyp}). Initially $\Gamma=\Gamma_0=1$, $f_0(1)=1$, and $g_0(1)$ is used as a control parameter, having the critical value $g_c=4.54169\cdots$. The red solid line is the critical trajectory ending at the critical fixed point $(0,1/3)$. The blue dotted lines and the green broken lines correspond to the inactive and active phase, respectively.}
\end{figure}

Unfortunately, the variation of the order parameters with $\Delta$ is not determined solely by the leading terms but also the next-to-leading one in $f_0$ is needed. Eq. (\ref{flow_g_simp}) implies the corrections to be of the form
$g_0=(\overline{a}-1)\Gamma^{-1}+g_1\Gamma^{-\gamma}\dots$ and $f_0=a+f_1\Gamma^{-\gamma-1}\dots$, with the constants fulfilling $\gamma=\frac{f_1}{g_1}\frac{1-a}{a^2}$ but otherwise leaving them unspecified. To fix these constants, the missing flow equation would be needed. An obvious problem with Eq. (\ref{flow_f_simp}) is that it enforces a wrong prefactor to the leading term of $g_0$, $g_0\simeq\overline{a}\Gamma^{-1}$.
We can naively correct this fault of Eq. (\ref{flow_f_simp}) by writing a hypothetical flow equation 
\be
f_0'=-f_0[g_0-(\overline{a}-1)\Gamma^{-1}]. \label{flow_f_hyp} 
\ee
This does not affect the critical scaling of order parameters with $\Gamma$, which is solely determined by Eq. (\ref{flow_g_simp}), but fixes the unknown constants in the correction terms and provides a prediction for the off-critical behavior of the order parameters. These, as we will see, are close to the estimates obtained by the numerical analysis of the NN2 scheme. In the sequel, we will therefore analyze the hypothetical flow equations.
The flow diagram constructed by numerical integration of these equations can be seen in Fig. \ref{fig_sefd}. 
The trajectories can be shown to be given by the equations
\be
\overline{a}f_0-\ln(\overline{a}f_0)=g_0\Gamma-(\overline{a}-1)\ln(g_0\Gamma)+{\rm const}.
\ee
Note that, as opposed to the trajectories of the PL model in Eq. \ref{tra}, these equations explicitly contain $\Gamma$.  
Using that, at the critical fixed point $f_0=0$ and $g_0\Gamma\to\overline{a}-1$, the critical trajectory is given by
$\overline{a}f_0-1-\ln(\overline{a}f_0)=g_0\Gamma-\overline{a}+1-(\overline{a}-1)\ln\frac{g_0\Gamma}{\overline{a}-1}$.

For the exponent $\gamma$ appearing in the next-to-leading term in $g_0(\Gamma)$ and $f_0(\Gamma)$ at criticality, we obtain
\be
\gamma=\sqrt{\overline{a}-1}. 
\ee
Thus, in general, the functions $g_0(\Gamma)$ and $f_0(\Gamma)$ are non-analytic functions of $1/\Gamma$ at the critical fixed point.  
The exponent $\gamma$ determines the shape of the critical trajectory near the fixed point through
\be 
f_0(\Gamma)-a\sim g_0^{\gamma}.
\label{se_shape}
\ee
The following considerations about the off-critical behavior of the order parameters are accurately supported by numerical analyses of the hypothetical flow equations. 
Let us first consider the persistence in the inactive phase ($\Delta<0$), which is proportional to the fixed point value $\frac{1}{z}$ of $g_0$. As can be seen in the flow diagram in Fig. \ref{fig_sefd}, a slightly off-critical trajectory with $|\Delta|\ll 1$ will stay close to the critical one up to some $g_0^*(\Delta)$, beyond which a crossover occurs and the trajectory breaks down rapidly. We may then assume that, up to the crossover, the vertical deviation from the critical trajectory remains essentially $O(|\Delta|)$, whereas in the subsequent part of the trajectory $g_0$ hardly changes and we may write $g_0^*(\Delta)\sim \frac{1}{z(\Delta)}$. Using the shape of the critical trajectory in Eq. (\ref{se_shape}), we obtain finally
\be 
\pi_s(\Delta)\sim \frac{1}{z(\Delta)}\sim |\Delta|^{\frac{1}{\gamma}},
\ee 
as $\Delta\to 0$. 

\begin{figure}[ht]
\includegraphics[width=8cm]{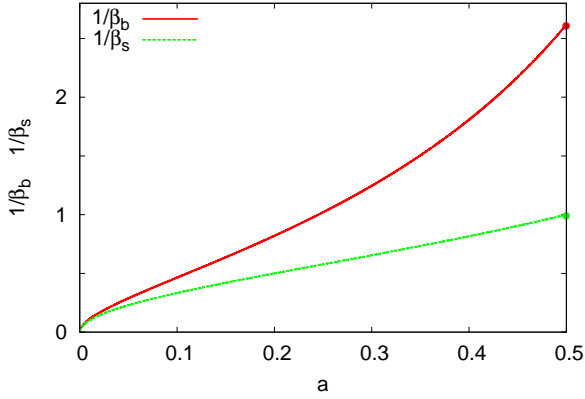}
\caption{\label{fig_beta} The reciprocal order-parameter exponents obtained by the hypothetical flow equations of the CP with SE dispersal kernel, plotted against $a$. The red and green dots indicate the order-parameter exponents $\beta_b=\frac{3-\sqrt{5}}{2}$ and $\beta_s=1$ of the short-range model, respectively \cite{hiv}.} 
\end{figure}

Next, let us consider the density order parameter in the active phase ($\Delta\gtrsim 0$). We assume again that there is a crossover scale $\Gamma^*(\Delta)$ within which the decrease of the order parameter essentially follows Eqs. (\ref{rho_se_surf}-\ref{rho_se_bulk}) valid at criticality, and beyond which it saturates to a finite limiting value. Furthermore, we assume that the crossover 
value $g_0^*(\Delta)\sim 1/\Gamma^*(\Delta)$ scales with $\Delta$ in the same way as below the critical point, $g_0^*(\Delta)\sim\Delta^{\frac{1}{\gamma}}$.
Using Eqs. (\ref{rho_se_surf}-\ref{rho_se_bulk}), this results in the dependence of the order parameter on $\Delta$ close to the critical point:    
\beqn
\rho_s(\Delta)&\sim& \Delta^{\beta_s} \label{rho_se_surf_2} \\
\rho_b(\Delta)&\sim& \Delta^{\beta_b}, \label{rho_se_bulk_2}
\eeqn
with the surface and bulk order-parameter exponents 
\beqn
\beta_s&=&\frac{x_s}{a\gamma}=\sqrt{\overline{a}-1} \\ 
\beta_b&=&\frac{x_b}{a\gamma}=\frac{2\overline{a}-1-\sqrt{4\overline{a}-3}}{2\sqrt{\overline{a}-1}},
\eeqn
respectively. The variation of these exponents with the parameter $a$ is shown in Fig. \ref{fig_beta}.  

\begin{figure}[ht]
\includegraphics[width=8cm]{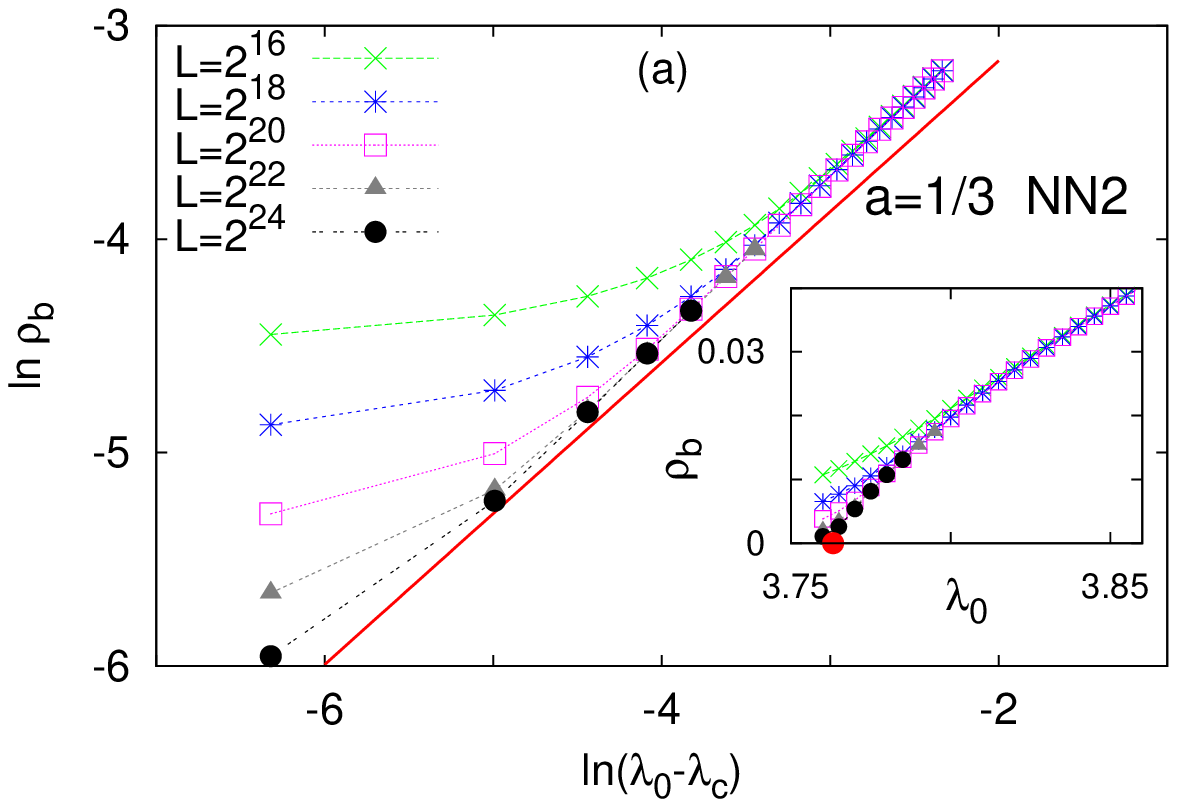}
\includegraphics[width=8cm]{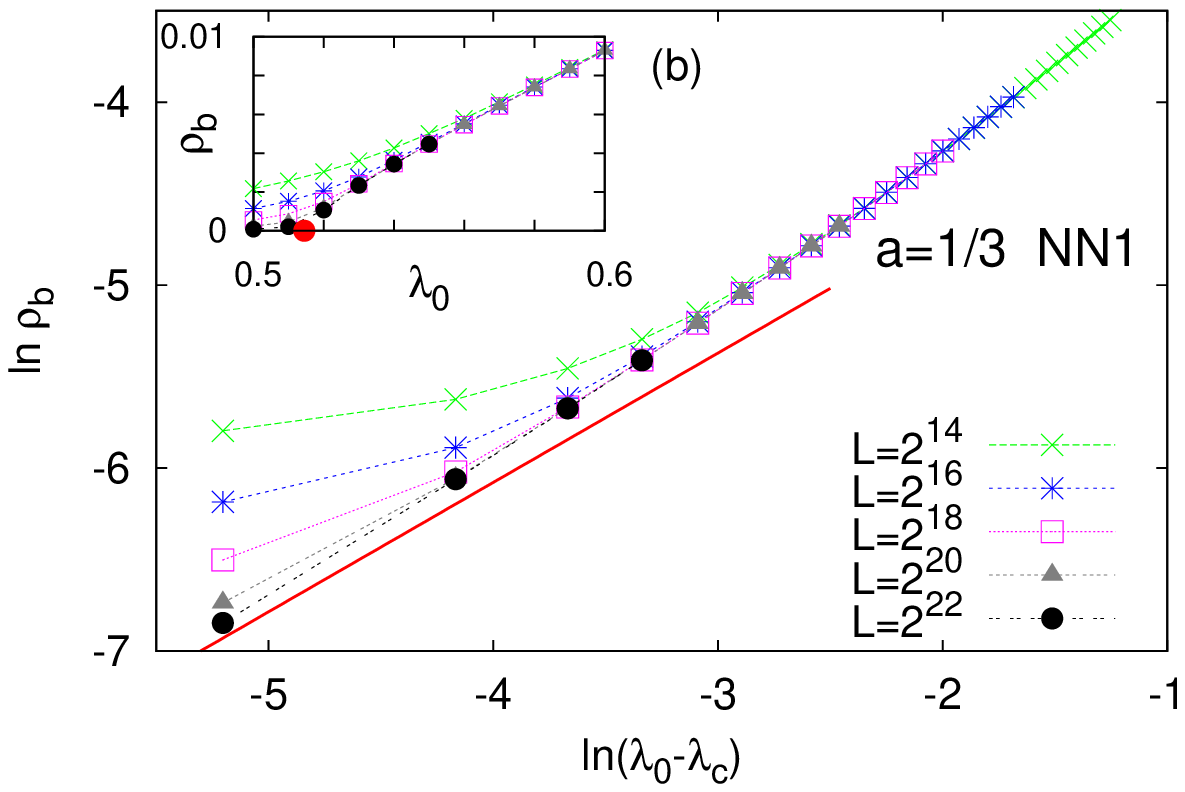}
\caption{\label{fig_opse3} 
The variation of the bulk density order parameter with the control parameter for the SE dispersal kernel with $a=1/3$. The data were obtained numerically by the NN2 scheme (a) and by the NN1 scheme (b). The solid line has a slope $\beta_b=\sqrt{2}/2=0.707\dots$ as predicted by the hypothetical flow equations. The red dot in the insets indicates the critical value of the control parameter, which is $\lambda_c=3.7632(5)$ (a) and $\lambda_c=0.5145(5)$ (b).}
\end{figure}

\begin{figure}[ht]
\includegraphics[width=8cm]{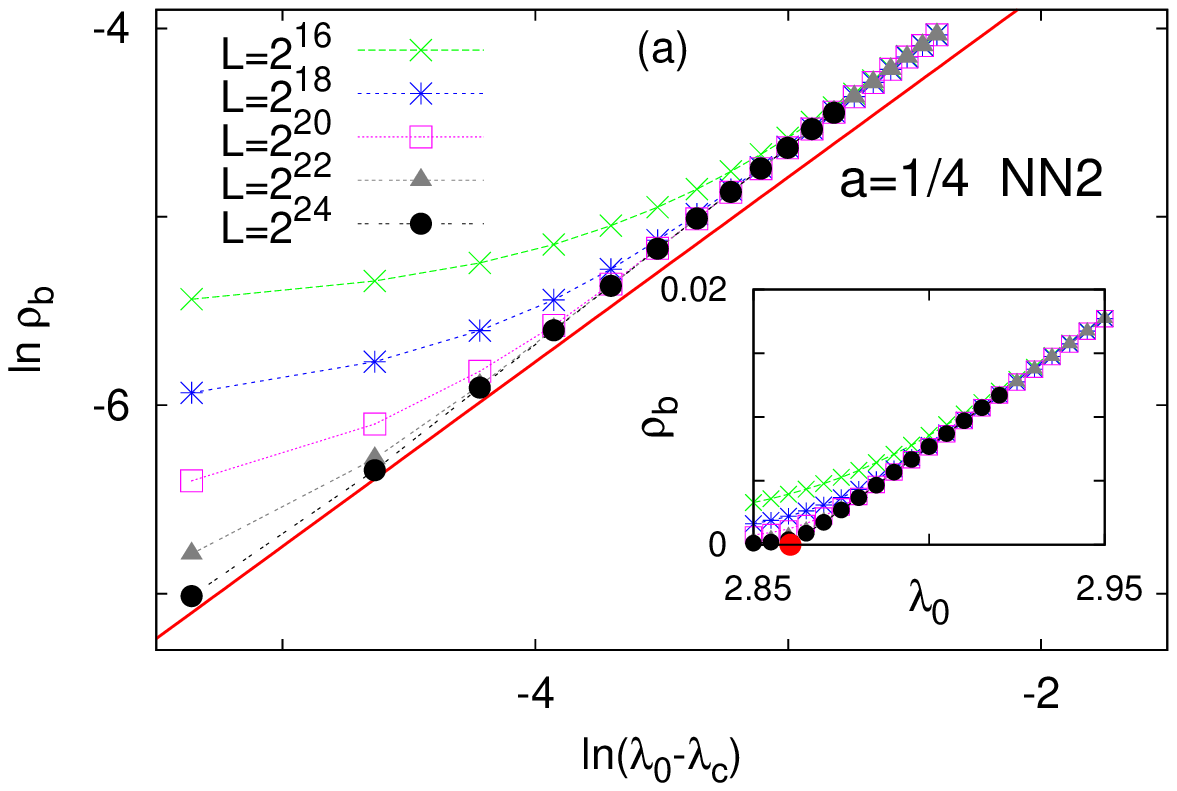}
\includegraphics[width=8cm]{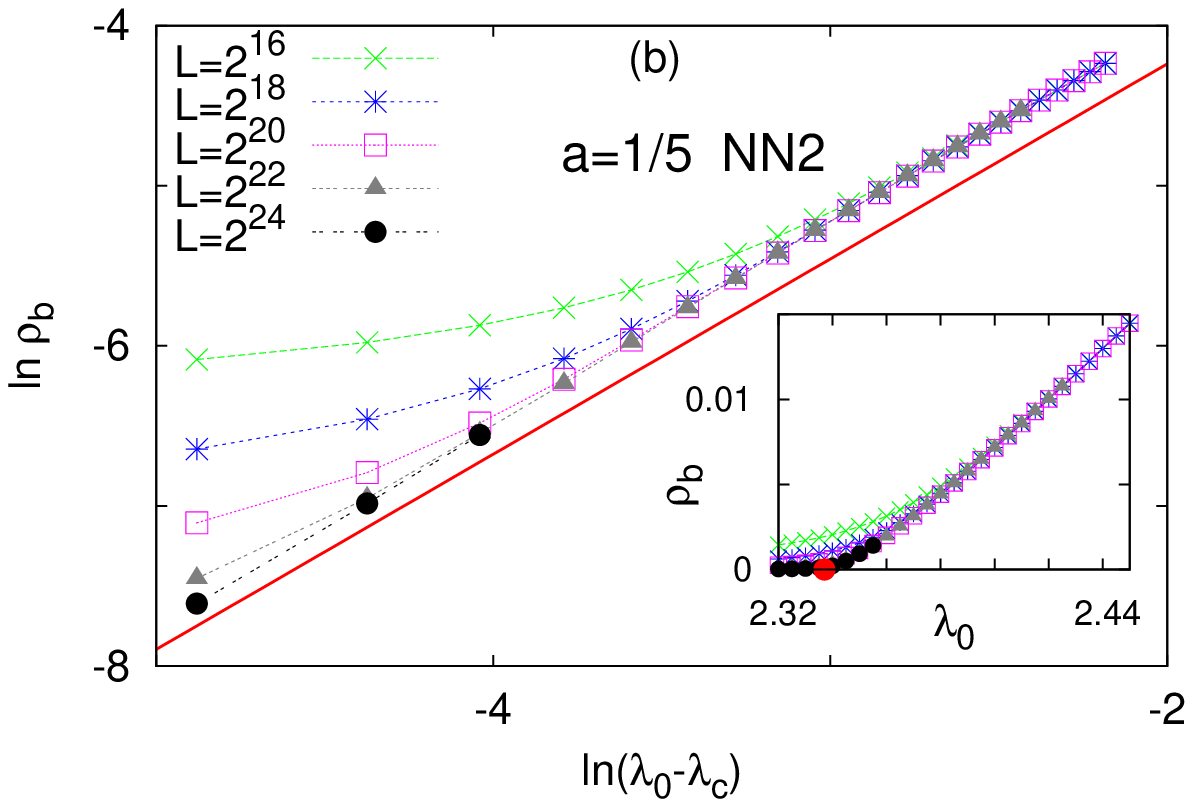}
\caption{\label{fig_opse45} 
The variation of the bulk density order parameter with the control parameter for the SE dispersal kernel with $a=1/4$ (a) and $a=1/5$ (b). The data were obtained numerically by the NN2 scheme. The solid line has a slope $\beta_b=\frac{7-\sqrt{13}}{2\sqrt{3}}=0.979\dots$ (a) and $\beta_b=\frac{9-\sqrt{17}}{4}=1.219\dots$ (b) as predicted by the hypothetical flow equations. The red dot in the insets indicates the critical value of the control parameter, which is $\lambda_c=2.8605(5)$ (a) and $\lambda_c=2.337(1)$ (b).}
\end{figure}

We have confronted the prediction of the hypothetical NN3 scheme about the vanishing of the bulk density order parameter with numerical analyses of the NN1 and NN2 schemes. 
The details of the numerical calculations with the NN1 scheme were the same as described for the PL dispersal kernel. For the NN2 scheme, the $\mu$ rates were drawn from a uniform distribution with the support $[0.1,1.1]$, while the initial distances between adjacent sites were uniformly distributed in $[0.9,1.1]$. 
The location of the critical point was determined by calculating the decimation ratio $r(L)$ at the last step of the renormalization and using that, at the critical point, $r(L)\to \frac{a}{1-a}$. The variation of the bulk density order parameter with the control parameter in the active phase is shown for $a=1/3,1/4$ and $1/5$ in Figs. \ref{fig_opse3}-\ref{fig_opse45}.    
As can be seen, the slope of data obtained by the NN2 scheme in the linearized plots, which is the bulk order-parameter exponent $\beta_b$, is close to the prediction in Eq. (\ref{rho_se_bulk_2}) obtained by the hypothetical flow equations, the relative differences being $4\%$.
Fig. \ref{fig_opse3}b shows that the long-range interactions between interior sites of neighboring clusters, which are taken into account in the NN1 scheme bring considerable corrections to the small-$\Delta$ behavior obtained by the NN2 scheme, appearing as a slow change of the local slopes with decreasing $\Delta$.

\subsection{The log-normal dispersal kernel}

\begin{figure}[ht]
\includegraphics[width=8cm]{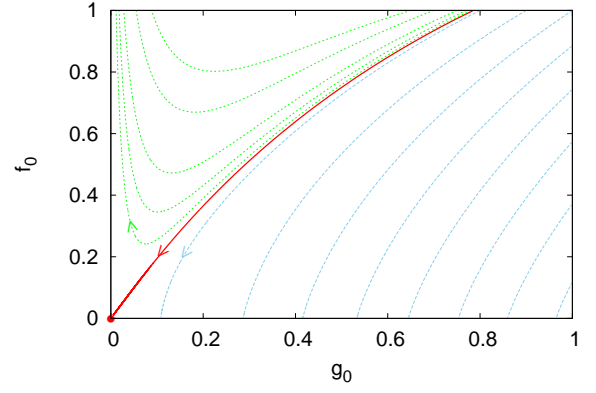}  
\caption{\label{fig_ln_flow} Flow diagram of the CP with a LN dispersal kernel ($a=2$, $b=1$) in the NN3 scheme.  Initially $\Gamma=\Gamma_0=1$, $f_0(1)=1$, and $g_0(1)$ is varied. The red solid line is the critical trajectory ending at the critical fixed point $(0,0)$. The blue dotted lines and the green broken lines correspond to the inactive and active phase, respectively.}
\end{figure}

The enhanced power-law dispersal kernel is given by 
\be 
\lambda(l)=\lambda_0e^{-b(\ln l)^{a}},
\ee
with constants $b>0$ and $a>1$. Here, the initial distribution of distances is restricted to the range $l\ge 1$. Note that $a=1$ is just the PL dispersal kernel, and $a=2$ represents the LN dispersal kernel.
The characteristic function is thus $\Theta(\Gamma)=B\Gamma^{\overline{a}}$, having a derivative 
\be 
\Theta'=B\overline{a}\Gamma^{\overline{a}-1},
\ee
where we introduced the constants $\overline{a}=1/a$ and $B\equiv b^{-\overline{a}}$. 
The flow diagram obtained by the numerical integration of Eqs. (\ref{flow_g_simp}) and (\ref{flow_f_simp}) for the LN dispersal kernel ($a=2$) is shown in Fig. \ref{fig_ln_flow}.   
For this model, we could not find the equations of trajectories. Nevertheless, the asymptotic dependence of $g_0$ and $f_0$ on $\Gamma$ along the critical trajectory can be determined: 
\beqn
g_0(\Gamma)&=&B\overline{a}\Gamma^{-1+\overline{a}}+\overline{a}\Gamma^{-1}+O(\Gamma^{-1-\overline{a}}),  \label{ln_g} \\
f_0(\Gamma)&=&\frac{1-\overline{a}}{B\overline{a}}\Gamma^{-\overline{a}}+B^{-2}\Gamma^{-2\overline{a}}+O(\Gamma^{-3\overline{a}}). \label{ln_f}
\eeqn
Thus, the decimation ratio in Eq. (\ref{r}) tends to zero according to 
\be 
r(\Gamma)\simeq \frac{a-1}{B}\Gamma^{-\overline{a}} 
\ee
as the critical fixed point is approached and the dynamical relationship, as it follows from Eq. (\ref{dyn}), is of the form
\be 
l\sim e^{B\Gamma^{\overline{a}}}\Gamma.
\ee
Substituting the asymptotic forms of $g_0$ and $f_0$ into Eqs. (\ref{u_simp}-\ref{v_simp}), we obtain for the surface and bulk density order parameter for large $\Gamma$ along the critical trajectory
\beqn
\rho_s(\Gamma)&\sim&e^{-B\Gamma^{\overline{a}}}\Gamma^{-\overline{a}}, \label{ln_rho_s_gamma} \\
\rho_b(\Gamma)&\sim&e^{-B\Gamma^{\overline{a}}+c\Gamma^{\overline{a}/2}},
 \label{ln_rho_b_gamma}
\eeqn
where $c=2\sqrt{B(a-1)}$, while for the surface persistence we have in leading order 
\be 
\pi_s(\Gamma)\sim g_0(\Gamma)\sim \Gamma^{-1+\overline{a}}.
\ee

\begin{figure}[ht]
\includegraphics[width=8cm]{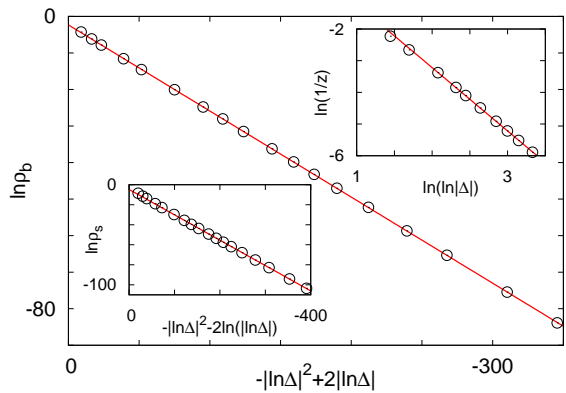}  
\caption{\label{fig_lnmpz} Variation of order parameters with the reduced control parameter obtained by the NN3 scheme of the LN model with $a=2$ and $b=1$. Initially 
$\Gamma=\Gamma_0=1$, $f_0(1)=1$, while $\Delta=g_c-g_0(1)$ is used as a reduced control parameter, where $g_c=0.785660168019\dots$ is the critical value of $g_0(1)$. 
The upper and lower insets show the surface persistence and density, in a linearized plot according to Eq. (\ref{ln_pers}) and Eq. (\ref{ln_rho_s_delta}), respectively. The slope of the straight line in the upper inset is $-2$. The main figure shows the bulk order parameter, linearized according to  Eq. (\ref{ln_rho_b_delta}).} 
\end{figure}

Next, we turn to the question how the order parameters depend on the control parameter close to the critical point, and we focus on the case of the LN dispersal kernel, $a=2$. As it turns out by a numerical analysis of the flow equations, the fixed-point value $1/z(\Delta)$ of $g_0$ in the inactive phase ($\Delta\lesssim 0$) does not related to $\Delta$ in a power-law fashion but the slightly off-critical trajectories are repelled much more strongly from the critical one. As it is shown in the upper inset of Fig. \ref{fig_lnmpz}, the numerically determined fixed-point values $1/z(\Delta)$ accurately follow the law
\be 
\frac{1}{z(\Delta)}\sim (\ln|\Delta|)^{-2}.
\label{ln_pers}
\ee 
Thus, the surface persistence vanishes in this phase according to $\pi_s\sim (\ln|\Delta|)^{-2}$ as $\Delta\to 0$. 

Concerning the density order parameter in active phase ($\Delta\gtrsim 0$), we assume again the existence of a crossover scale $\Gamma^*$ within which the flow is almost critical and beyond which the order parameter essentially saturates. Furthermore, we assume that the corresponding crossover parameter $g_0^*\sim [\Gamma^*]^{-1+\overline{a}}$ deviates from the critical curve in the same way as in the inactive phase, i.e. $g_0^*\sim (\ln\Delta)^{-2}$. For $a=2$, we have then $\Gamma^*(\Delta)\sim (\ln\Delta)^4$, which can be substituted into Eqs. (\ref{ln_rho_s_gamma}-\ref{ln_rho_b_gamma}) to yield for the variation of surface and bulk density order parameters with $\Delta$
\beqn
\rho_s(\Delta)&\sim&e^{-B(\ln\Delta)^{2}-2\ln|\ln\Delta|}, 
\label{ln_rho_s_delta} \\
\rho_b(\Delta)&\sim&e^{-B(\ln\Delta)^{2}+c|\ln\Delta|},
\label{ln_rho_b_delta}
\eeqn
respectively, as $\Delta\to 0$. 
The slightly off-critical order parameters obtained by the numerical integration of the flow equations are in agreement with these asymptotic forms, as shown in Fig. \ref{fig_lnmpz}.

\section{Discussion}
\label{discussion}

We have studied in this paper disordered one-dimensional contact processes with heavy-tailed dispersal, focusing on the behavior of order parameters near the critical point. 
We formulated a general renormalization framework valid for the dispersal kernels used in ecological studies, and after a series of approximations we obtained an analytically tractable asymptotic theory. 
The functional forms of the variation of different order parameters with the control parameter found in this work, as well as their dynamical scaling at the critical point, partially known from earlier works \cite{2dlr,stretched,lrpers}, are summarized in Table \ref{table}.

\begin{table*}[ht]
\begin{center}
\begin{tabular}{|c||c|c|c|}
\hline dispersal kernel & stretched exponential (SE)  & log-normal (LN)  & power law (PL) \\
\hline
\hline $\lambda(l)$ & $\lambda_0e^{-bl^a}$ ($a<\frac{1}{2}$)& $\lambda_0e^{-b(\ln l)^2}$ & $\lambda_0l^{-\alpha}$ ($\alpha>\frac{3}{2}$)\\
\hline critical fixed point & IDFP & IDFP & FDFP \\
\hline dynamical scaling & $l\sim(\ln\frac{t}{t_0})^{1/a}$  & $l\sim e^{B(\ln\frac{t}{t_0})^{1/2}}\ln\frac{t}{t_0}$ & $l\sim (\frac{t}{t_0})^{1/\alpha}(\ln\frac{t}{t_0})^2$ \\
\hline  $\rho_b(t)$  & $(\ln\frac{t}{t_0})^{-x_b/a}$ & $e^{-B(\ln\frac{t}{t_0})^{1/2}+c(\ln\frac{t}{t_0})^{1/4}}$  &  $(\frac{t}{t_0})^{-1/\alpha}$ \\
\hline  $\rho_s(t)$  & $(\ln\frac{t}{t_0})^{-x_s/a}$  & $e^{-B(\ln\frac{t}{t_0})^{1/2}}(\ln\frac{t}{t_0})^{-1/2}$ & $(\frac{t}{t_0})^{-1/\alpha}(\ln\frac{t}{t_0})^{-2}$ \\
\hline  $\pi_s(t)$  & $(\ln\frac{t}{t_0})^{-1}$ & $(\ln\frac{t}{t_0})^{-1/2}$ & $\frac{1}{\alpha}+2(\ln\frac{t}{t_0})^{-1}$ \\
\hline $\rho_b(\Delta)$ & $\Delta^{\beta_b}$ & $e^{-B(\ln\Delta)^{2}+c|\ln\Delta|}$ & $e^{-c/\sqrt{\Delta}}$ \\
\hline $\rho_s(\Delta)$ & $\Delta^{\beta_s}$ & $e^{-B(\ln\Delta)^{2}-2\ln|\ln\Delta|}$ & $\Delta e^{-c/\sqrt{\Delta}}$ \\
\hline $\pi_s(\Delta)$  & $(-\Delta)^{\frac{1}{\gamma}}$ & $[\ln(-\Delta)]^{-2}$ & $\frac{1}{\alpha}+O[(-\Delta)^{1/2}]$ \\
\hline
\end{tabular}
\end{center}
\caption{\label{table} Summary of the asymptotic scaling behavior of the bulk density $\rho_b$, the surface density $\rho_s$, and the surface persistence $\pi_s$ for three different dispersal kernels. The dynamical relationship between the length scale $l$ and time scale $t$ at the critical point is also shown. $\rho_b(t)$, $\rho_s(t)$, and  $\pi_s(t)$ denote the asymptotic time-dependence at the critical point, which, for the PL and the SE model, has also been derived earlier in Refs. \cite{2dlr,stretched,lrpers}. $\rho_b(\Delta)$, $\rho_s(\Delta)$, and $\pi_s(\Delta)$ denote the off-critical behavior in the limit $\Delta\to 0^+$ (active phase) for the densities and $\Delta\to 0^-$ (inactive phase) for the persistence. The constants appearing here are defined in the text.}
\end{table*}
Concerning the density order parameter, our findings can be summarized qualitatively as follows: the broader is the dispersal the faster is the vanishing of the density as the critical point is approached. For the most rapidly decaying dispersal kernel, the stretched exponential one, $\lambda(l)\sim e^{-bl^a}$, the density vanishes algebraically with the control parameter, $\rho(\Delta)\sim \Delta^{\beta}$. Decreasing $a$, i.e. making the dispersal kernel broader and broader, the order-parameter exponent $\beta$ increases monotonically, starting from its short-range value at $a=1/2$ to infinity, thus the vanishing of the density becomes less and less singular. For the log-normal distribution kernel, which decays more slowly than any SE function, the density vanishes as an enhanced power law with $\Delta$, thus the order-parameter exponent is formally infinite. 
For the most slowly decaying PL dispersal kernel, the density vanishes even more rapidly, following an exponential function of $1/\sqrt{\Delta}$, again with an infinite $\beta$ exponent.     

Besides the density, we also considered the persistence which can be regarded as an order parameter becoming non-zero in the inactive phase. 
We have found that the broadening of the dispersal kernel has an opposite effect on the persistence compared to the density: it makes the vanishing of the persistence sharper. For the SE dispersal kernel, the persistence vanishes algebraically, and decreasing $a$, i.e. broadening the dispersal, the order-parameter exponent decreases monotonically from its short-range value at $a=1/2$ toward zero. For the LN dispersal kernel, the persistence exhibits a logarithmic singularity with a formally zero order-parameter exponent. Finally, for the even more slowly decaying PL dispersal kernel, the persistence will have a discontinuity at the critical point.    

As we already noted, the validity of the SDRG method for arbitrarily weak disorder is not rigorously known for the nearest-neighbor CP \cite{wd_footnote}.
This problem is also inherited by the CP with heavy-tailed dispersal studied in this paper. We stress that even in the range of validity of the SDRG approach, the results obtained by method are expected to be valid only asymptotically, beyond a disorder-dependent crossover scale which increases with a decreasing strength of disorder.

As mentioned in the Introduction, spatially extended ecological systems are frequently affected by an environmental gradient, which can be modeled by a linear variation of the local control parameter with the position in some direction. In such systems at low gradients, the local density is essentially determined by the local control parameter, thus the dependence of the density in a gradient free system on the control parameter (studied in this paper) appears here as a variation of the local density with the coordinate along the gradient direction. In this context, our results imply that a broader dispersal leads to a faster decline of the local density with the position along the direction of the gradient.   

It is worth mentioning that the renormalization theory presented in this paper also describes the zero temperature quantum phase transition of random transverse-field Ising chains with ferromagnetic long-range couplings of strength $\lambda(l)$ \cite{jki}. The order parameters $\rho_b$ and $\rho_s$ analyzed here correspond in that model to the bulk and surface magnetization, respectively.  

As further directions of this research, it would be desirable to find the missing elements in the analytic description of the model with SE and LN dispersal kernel, to extend the investigations to more realistic two-dimensional systems and to confront the predictions of the renormalization theory about the order parameter scaling obtained in this work with Monte Carlo simulations. These are left for future research.

\begin{acknowledgments}
The author thanks F. Igl\'oi, B. Oborny, G. Ro\'osz, and G. \'Odor for useful discussions. This work was supported by the National Research, Development and Innovation Office NKFIH under Grant No. K128989.
\end{acknowledgments}

\appendix
\section{Master equation for rates}
\label{app:master}

Staying within the NN3 scheme, let us denote the distribution of rates at scale $\Omega$ by $P_{\Omega}(\mu)$ and $R_{\Omega}(\lambda)$. We assume that the scale is infinitesimally shifted from $\Omega$ to $\Omega+d\Omega$ and write down how the distribution $P_{\Omega}(\mu)$ changes.  
Let us first consider the change of the normalization of $P_{\Omega}(\mu)$. Due to the shrinking of the support, it decreases by $P_{\Omega}(\Omega)|d\Omega|$. In addition to this, it is also affected by $\lambda$-decimations occurring with a probability $R_{\Omega}(\Omega)|d\Omega|$. By such an event, two $\mu$ rates are eliminated and a new one is generated, thus the net balance is $-1$. If, however, the generated $\tilde\mu$ is greater than $\Omega$, i.e. an anomalous $\lambda$-decimation event occurs, it will we be immediately eliminated by a subsequent $\mu$-decimation, resulting in a net balance of $-2$. Denoting the probability of normal $\lambda$-decimation events by $s_{\Omega}={\rm Prob}(\tilde\mu<\Omega)$, the normalization changes in total to 
$1-P_{\Omega}(\Omega)|d\Omega|-R_{\Omega}(\Omega)|d\Omega|[s_{\Omega}+2(1-s_{\Omega})]$.   
For the distribution of $\mu$, we can write 
\beqn
P_{\Omega+d\Omega}(\mu)=\left\{P_{\Omega}(\mu) + R_{\Omega}(\Omega)|d\Omega|
[-2P_{\Omega}(\mu)+I(\mu)]\right\}\times \nonumber \\
\times\frac{1}{1-P_{\Omega}(\Omega)|d\Omega|-R_{\Omega}(\Omega)|d\Omega|(2-s_{\Omega})}. \nonumber  \\
\eeqn
Here, the first term in the brackets on the r.h.s. describes the loss of eliminated $\mu$ rates, the second term, $I(\mu)$ denotes the distribution of generated ones, and the last factor restores the normalization of $P_{\Omega+d\Omega}(\mu)$. This leads to the differential equation
\be
\frac{\partial P_{\Omega}(\mu)}{\partial\Omega}=  [R_{\Omega}(\Omega)s_{\Omega}-P_{\Omega}(\Omega)]P_{\Omega}(\mu) - R_{\Omega}(\Omega)I(\mu).
\ee
Now we reformulate this equation by using $\Gamma$ instead of $\Omega$, and the reduced variables, $\beta$ and $\zeta$, the distributions of which are denoted by $g_{\Gamma}(\beta)$ and $f_{\Gamma}(\zeta)$, respectively. As can be seen from Eq. (\ref{beta_rule}), $I(\mu)$ becomes a convolution in terms of $\beta$ and we  are lead in a straightforward way to Eq. (\ref{master_g}).  

Next, we formulate a master equation for $f_{\Gamma}(\zeta)$, starting from the distribution $F_{\Gamma}(l)$ of distance variables $l$. This changes by $\mu$-decimations, in which two $l$ variables are deleted and a new one is generated, as well as by anomalous $\lambda$-decimations, which are followed by an immediate $\mu$-decimation. This latter decimation will be slightly different from $\mu$-decimations in that the generated new distance variable in total will contain also the length variable $\lambda^{-1}(\Omega)$ decimated in the first part of the anomalous $\lambda$-decimation, $\tilde l=l_{n-1,n}+l_{n,n+1}+\lambda^{-1}(\Omega)$. In terms of $\zeta$, we have then $\tilde\zeta=\zeta_{n-1,n}+\zeta_{n,n+1}+2$, thus the additive constant is $2$ instead of $1$. Nevertheless, as we argued in the main text, in the domain of validity of the NN3 scheme, the constant terms can be neglected, therefore the difference between the two types of decimations is irrelevant.   
When the logarithmic rate scale is shifted from $\Gamma$ to $\Gamma+d\Gamma$, the distribution $F_{\Gamma}(l)$ changes as follows:
\begin{widetext}
\be
F_{\Gamma+d\Gamma}(l)=
\left\{F_{\Gamma}(l)+d\Gamma[g_0 + f_0\Theta'(1-p_{\Gamma})][-2F_{\Gamma}(l)+(F_{\Gamma}\ast F_{\Gamma})(l)]\right\}\frac{1}{1-[f_0\Theta'(2-p_{\Gamma})+g_0]d\Gamma}.
\label{F}
\ee
\end{widetext}
Here, the second term on the r.h.s. describes the elimination of two $l$ variables and a generation of a new one, having a distribution $(F_{\Gamma}\ast F_{\Gamma})(l)=\int_{\lambda^{-1}(\Omega)}^l F_{\Gamma}(l')F_{\Gamma}(l-l')dl'$. Such events occur by $\mu$-decimations and by anomalous $\lambda$-decimations with probabilities $g_0d\Gamma$ and $f_0\Theta'(1-p_{\Gamma})d\Gamma$, respectively. The last factor is again for keeping the distribution normalized. Eq. (\ref{F}) can be recast as a differential equation  
\beqn
\frac{\partial F_{\Gamma}(l)}{\partial\Gamma}=
F_{\Gamma}(l)[f_0\Theta'p_{\Gamma}-g_0] + \nonumber \\
+ [f_0\Theta'(1-p_{\Gamma})+g_0](F_{\Gamma}\ast F_{\Gamma})(l). \nonumber
\eeqn
Rewriting this equation in terms of the reduced variable $\zeta$, we arrive ultimately at Eq. (\ref{master_f}).

\section{Master equation for order parameters}
\label{app:order}

\subsection{Density} 

Let us define $\sigma^{(i)}_{\Gamma}(\mu)d\mu$ as the probability that a given bulk ($i=b$) or surface ($i=s$) site has survived the $\mu$-decimations up to scale $\Gamma$ and it is part of a cluster having a deactivation rate in the range $[\mu,\mu+d\mu]$. When $\Gamma$ is shifted to $\Gamma+d\Gamma$, it will change due to $\lambda$-decimations on two sides (one side) of the containing cluster for bulk (surface) sites. The probability of $\lambda$-decimations is $n_if_0\Theta'd\Gamma$ with $n_s=1$, $n_b=2$, so we can write 
\be  
\sigma^{(i)}_{\Gamma+d\Gamma}(\mu)=\sigma^{(i)}_{\Gamma}(\mu) + n_if_0\Theta'd\Gamma[-\sigma^{(i)}_{\Gamma}(\mu) + I(\mu)],
\label{sigma_shift}
\ee
where the first term in the brackets is the loss term while the second one is the gain given by
$I(\mu)=\int\int\sigma^{(i)}_{\Gamma}(\mu_1) P_{\Omega}(\mu_2)\delta(\mu-\kappa\frac{\mu_1\mu_2}{\Omega})d\mu_1d\mu_2$.
Note that, as opposed to $P_{\Omega}(\Omega)$ and $R_{\Omega}(\Omega)$, $\sigma^{(i)}_{\Gamma}(\mu)$ is not normalized to one [the norm being $S_i(\Gamma)$], therefore no compensation factor needs to be included in Eq. (\ref{sigma_shift}).   
Moreover, no further care has to be taken of anomalous $\lambda$-decimations (with $\tilde\mu>\Omega$) since the generated $\tilde\mu$ is in this case outside of the support of $\sigma^{(i)}_{\Gamma}(\mu)$ and is thus automatically put down to the losses of $S_i(\Gamma)$.  
Eq. (\ref{sigma_shift}) can be recast as the differential equation
\be
\frac{\partial\sigma^{(i)}_{\Gamma}(\mu)}{\partial\Gamma} = 
-n_if_0\Theta'[\sigma^{(i)}_{\Gamma}(\mu) - I(\mu)].
\label{sigma_diff}
\ee
Rewriting this in terms of $s_{\Gamma}^{(i)}(\beta)$, we obtain Eq. (\ref{master_s}).

\subsection{Persistence}

We define $\mathcal{Q}_{\Gamma}(l)dl$ as the probability that the bond next to the first site of a semi-infinite system has not been eliminated by a $\lambda$-decimation up to scale $\Gamma$ and its length variable lies in the range $[l,l+dl]$. 
When $\Gamma$ is shifted to $\Gamma+d\Gamma$, $\mathcal{Q}_{\Gamma}(l)$ changes by $\mu$-decimations of the second site which have a probability $g_0d\Gamma$, as well as by anomalous $\lambda$-decimations of the second bond occurring with a probability $f_0\Theta'(1-p_{\Gamma})d\Gamma$. We can then write down the following differential equation
\beqn
\frac{\partial\mathcal{Q}_{\Gamma}(l)}{\partial\Gamma}=-[g_0+f_0\Theta'(1-p_{\Gamma})]
\times \nonumber \\ 
\times\left[\mathcal{Q}_{\Gamma}(l)-\int_{\lambda^{-1}(\Omega)}^l\mathcal{Q}_{\Gamma}(l')F_{\Omega}(l-l')dl'\right]. \nonumber \\
\eeqn
Using the reduced variable $\zeta$ instead of $l$, this can be reformulated as Eq. (\ref{master_q}).



\begin{thebibliography}{99}
\bibitem{wd_footnote} The question whether the IDFP is attractive for arbitrarily weak disorder \cite{hoyos_wd} or there exists a line of fixed points in the weak disorder regime \cite{hiv,nft} is at present undecided. Nevertheless, large-scale Monte Carlo simulations do not show indications of the latter scenario \cite{vd}.
\end{thebibliography}
\end{document}